\documentclass{PoS}

\title{INTEGRAL/IBIS and optical observations of the Crab nebula/pulsar polarisation}

\ShortTitle{INTEGRAL/IBIS and optical observations of the Crab nebula/pulsar polarisation}

\author{\speaker{Paul Moran} \\
        Centre for Astronomy, NUI Galway, Ireland\\
        E-mail: \email{p.moran4@nuigalway.ie}}

\author{Andy Shearer\\
        Centre for Astronomy, NUI Galway, Ireland\\
       E-mail: \email{andy.shearer@nuigalway.ie}}

\author{Christian Gouiff\`es\\
        CEA, IRFU, Service d'Astrophysique, 91191 Gif-sur-Yvette, France\\
        AIM, CEA/CNRS/Universit\'e Paris Diderot, SAp, Saclay, 91191 Gif-sur-Yvette, France\\
       E-mail: \email{christian.gouiffes@cea.fr}}

\author{Philippe Laurent\\
        CEA, IRFU, Service d'Astrophysique, 91191 Gif-sur-Yvette, France\\
        APC, 10 rue Alice Domont et Leonie Duquet, 75205 Paris Cedex 13, France\\
       E-mail: \email{philippe.laurent@cea.fr}}

\abstract{Previous INTEGRAL/IBIS observations have shown that the gamma-ray radiation of the Crab Nebula is highly polarised and remarkably aligned along the axis of rotation of the pulsar \cite{Forot08}. Their study was based on the first four years of operation of the satellite. Here we present an analysis based upon nearly ten years of operation. This new analysis allows a better characterisation of the polarisation fraction and position angle, as well as a measure of spectral energy distribution of the polarised component. These results are then compared to the known optical polarisation of the pulsar and nearby synchrotron knot. In the future we shall compare the gamma-ray polarisation with the phase-resolved optical polarisation using instruments such as GASP \cite{Kyne10}.}

\FullConference{An INTEGRAL view of the high-energy sky (the first 10 years) -
9th INTEGRAL Workshop and celebration of the 10th anniversary of the launch,\\
		October 15-19, 2012\\
		Biblioth\`eque Nationale de France, Paris, France}

\begin{document}

\section{Introduction}

Polarisation measurements of pulsars provide an unique insight into the geometry of their emission region, and therefore observational constraints on the theoretical models of the emission mechanism. From an understanding of the emission geometry, one can limit the competing models for pulsar emission, and hence understand how pulsars work - a problem which has eluded astronomers for almost 50 years. Here we present the preliminary results of a multi-wavelength campaign, whereby we study the polarisation of the Crab Nebula and pulsar using both the INTEGRAL/IBIS Compton mode (gamma-ray) and archival HST/ACS data (optical).

\section{INTEGRAL/IBIS Polarisation via Compton Mode}

The work presented here is a continuation of the work of \cite{Forot08}. Those authors studied the polarisation of the Crab Nebula and pulsar using the first four years of operation of the INTEGRAL satellite. We use the same procedure and software for our analysis of the data from 2007 September to 2012 April. The polarisation of celestial sources is measured using the IBIS Compton mode.
Photons entering IBIS (Imager on Board the INTEGRAL Satellite) are Compton scattered in ISGRI (Integral Soft Gamma-Ray Imager), the first detector plane, at a polar angle $\rm \theta$ from their incident direction and at an azimuth $\rm \psi$ from their incident electric vector. The photons are then absorbed in the second detector, PICsIT (Pixellated Ceasium Iodide Telescope). Hence, the polarisation of sources can be measured, since the scattering azimuth is related to the polarisation direction. Events recorded in both detectors within the same time window of 3.8 $\rm \mu$s are tagged as Compton events. However, not all of these events result from Compton scattering. Chance  coincidences can occur between events independently coming from the source, the sky, or the instrumental background. The majority of the Compton tagged events are due to the background events, which are removed using the shadowgram deconvolution. IBIS is an example of a coded-aperture Compton telescope. The design provides high-energy response, low background, and a wide field-of-view. Moreover, utilising the imaging properties of the coded mask, a high angular resolution and background subtraction is achieved (see Figure 1). For the timing analysis, the times of arrival of the Compton events are referred to the solar system barycentre and phase-folded using an ephemeris from Jodrell Bank. We have used the pulsar phase intervals of \cite{Kuiper01}: P1 (main pulse), P2 (inter pulse), B (Bridge emission), and OP (off-pulse emission) (see Figure 2). For more information about the analysis see \cite{Forot07} and the paper by Laurent et al. in these proceedings.

\section{Optical Polarisation Studies}

Strong polarisation is expected when the pulsar optical emission is generated by synchrotron radiation. \cite{Shklovsky53} suggested that the continuous optical radiation from the Crab Nebula was due to synchrotron radiation. This was later confirmed by \cite{Dombrovsky54} and \cite{Vashakidze54} who found that the optical radiation was polarised. Archival HST/ACS polarisation science frames of the Crab Nebula were obtained from the Mikulski Archive for Space Telescopes (MAST). The data comprises of a series of observations of the nebula with the HST/ACS taken in three different polarisers (0$^{\circ}$, 60$^{\circ}$  \& 120$^{\circ}$) between 2003 August and 2005 December. The raw images, which had already been flat-fielded, were geometrically aligned, combined and averaged with cosmic-ray removal using IRAF. For each set of observations, the images were analysed by the IMPOL software \cite{Walsh99}, which produces polarisation vector maps. In order to determine the polarimetry, aperture photometry was performed on the pulsar and synchrotron knot in each image using the IRAF task \textit{phot} (see ACS Data Handbook \cite{Pavlovsky04}). This work is intended to accurately map the polarisation of the Crab Nebula, and act as a guideline for future time-resolved polarisation measurements of the Crab pulsar using the Galway Astronomical Stokes Polarimeter (GASP). This is an ultra-high-speed, full Stokes, astronomical imaging polarimeter based on the Division of Amplitude Polarimeter (DOAP). It has been designed to resolve extremely rapid variations in objects such as optical pulsars and magnetic cataclysmic variables.

\section{Conclusion}

We have measured the polarisation of the Crab Nebula and pulsar using the INTEGRAL/IBIS Compton mode. Our results are consistent with those of \cite{Forot08}. They found that the off-pulse emission is highly polarised, and that the polarisation is aligned with the axis of rotation of the pulsar ($124\pm0.1^{\circ}$) \cite{Ng04}. We have also examined the phase-averaged optical polarisation of the Crab pulsar and its synchrotron knot using archival HST/ACS data. The results of the optical polarimetry of the Crab pulsar are in good agreement with those of \cite{Aga09} using OPTIMA, \cite{Wampler69}, and \cite{Kristen70}. We see that the knot is strongly polarised and we suggest that it is responsible for the highly polarised off-pulse emission. Our measurement of the polarisation PA of the synchrotron knot, $\rm PA=126.86\pm0.23^{\circ}$, agrees with the Crab torus $\rm PA=126.31\pm0.03^{\circ}$ \cite{Ng04}. We also found evidence for an apparent alignment between the pulsar polarisation PA (105.97$\pm2.00^{\circ}$) and proper motion vector (\cite{Kaplan08}; $290\pm2\pm9^{\circ}$ or $110\pm2\pm9^{\circ}$) (see Figure 4). Looking at the gamma-ray data, we again see evidence of an alignment between the polarisation position angle of the pulsar and it's rotation axis. Furthermore, we have compared our results to those obtained at other wavelengths (see Table 1). 

The optical polarisation maps show the variation of the polarisation throughout the inner nebula and particularly in the vicinity of the pulsar itself. One can distinctly see the overall structure of the inner nebula, the degree of polarisation of the knots and the synchrotron emission (see Figures 3 \& 4). The first optical polarisation maps of the Crab Nebula were those of \cite{Oort56}, \cite{Hiltner57} and \cite{Woltjer57}. 

In November 2012 we used GASP to measure both the linear and circular optical polarisation from the Crab pulsar, on time-scales of $\approx1$ millisecond. The analysis of these observations are still ongoing. Polarisation measurements give an unique insight into the geometry of the pulsar emission region (see \cite{McDonald11}). More multi-wavelength polarisation observations of pulsars will help to provide the much needed data to constrain the theoretical models, and hence solve the emission-model problem.

\section*{Acknowledgments} 

Some of the data presented in this paper were obtained from the Mikulski Archive for Space Telescopes (MAST). STScI is operated by the Association of Universities for Research in Astronomy, Inc., under NASA contract NAS5-26555. Support for MAST for non-HST data is provided by the NASA Office of Space Science via grant NNX09AF08G and by other grants and contracts. We thank Jeremy Walsh, ESO, for the use of his polarimetry software IMPOL to produce the polarisation maps. PM is grateful for his PhD funding from the Irish Research Council, and also thanks the French Embassy in Ireland for helping to fund this research project.

\begin{figure}[h]
\begin{center}
\includegraphics[width=60mm]{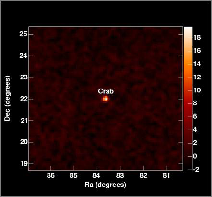}
\caption{Deconvolved significance map of the Crab Nebula and pulsar using the Compton mode, 200--800 keV, 1 Ms.}
\label{figure1}
\end{center}
\end{figure}

\begin{figure}[h]
\begin{center}
\includegraphics[width=75mm]{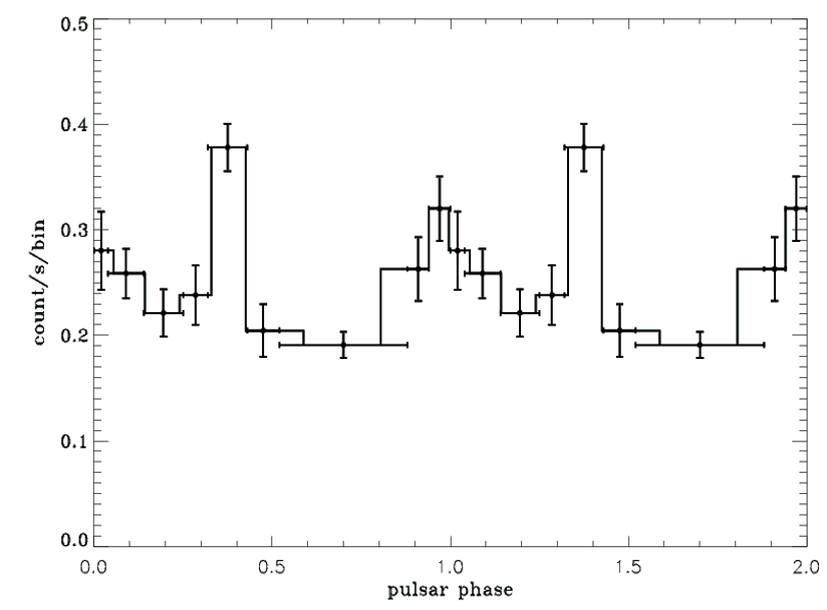}
\includegraphics[width=75mm]{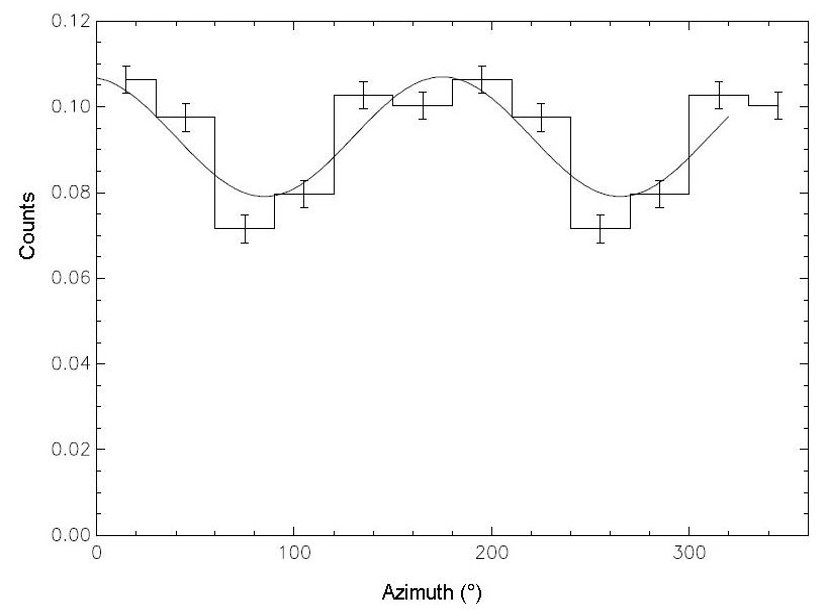}
\caption{Left: Compton mode lightcurve of the Crab pulsar, 200--600 keV, 2.6 Ms. Right: Azimuthal profile of the Crab Nebula and pulsar, 200--800 keV, 2.6 Ms.}
\label{figure2}
\end{center}
\end{figure}

\begin{figure}[h]
\begin{center}
\includegraphics[width=70mm]{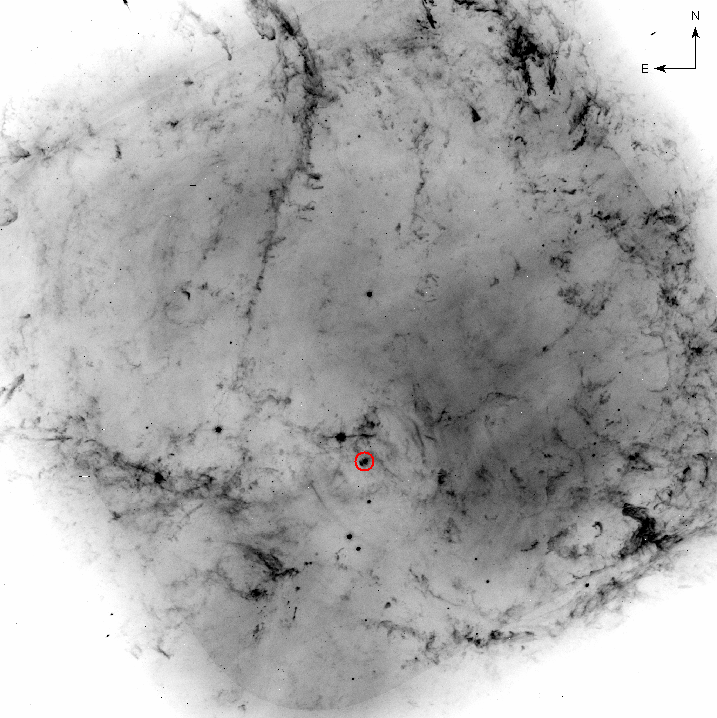}
\includegraphics[width=70mm]{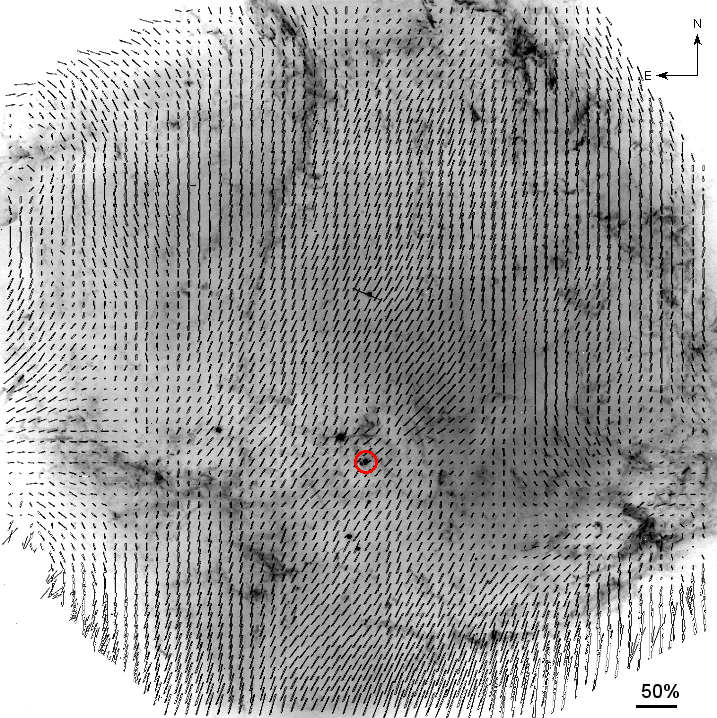}
\caption{Left: HST/ACS image of the Crab Nebula (1.76$^{\prime}\times$1.7$^{\prime}$, 2005 Sep 06). Right: Polarisation vector map of the Crab Nebula superimposed on the nebula. The legend shows the vector magnitude for 50\% polarisation. The location of the pulsar and knot is marked by the red circle.}
\label{figure3}
\end{center}
\end{figure}

\begin{figure}[h]
\begin{center}
\includegraphics[height=60mm]{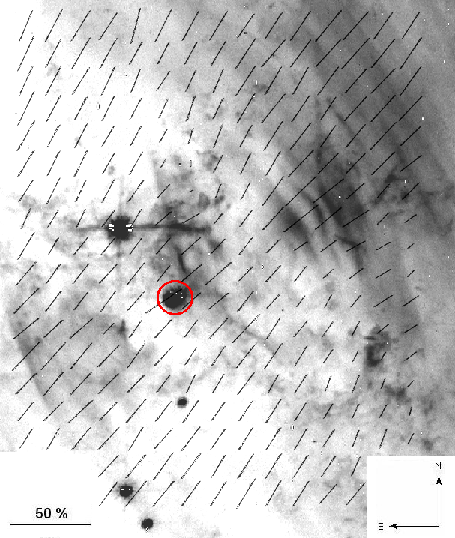}
\includegraphics[height=60mm]{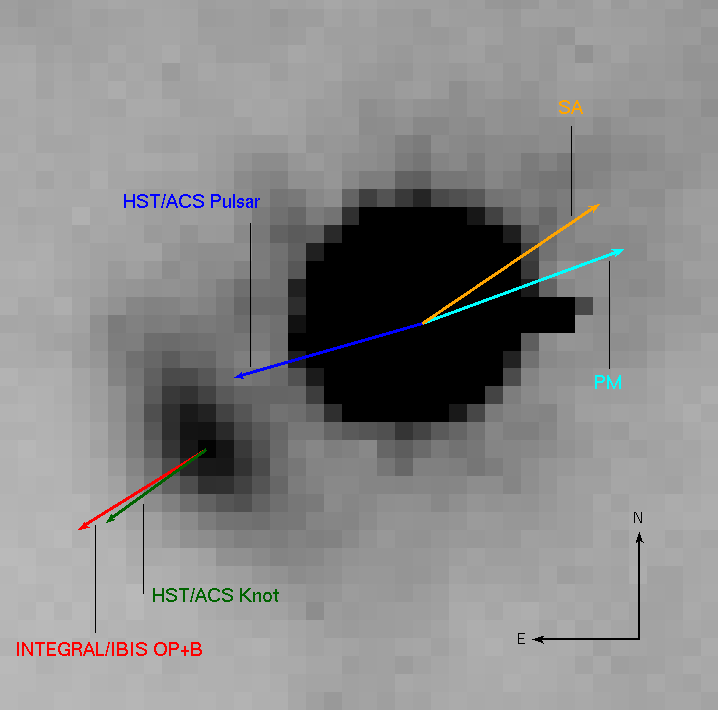}
\caption{Left: Polarisation vector map of the vicinity of the Crab pulsar and wisp region superimposed on the nebula. The legend shows the vector magnitude for 50\% polarisation. The location of the pulsar and knot is marked by the red circle. Right: Close up of the Crab pulsar region with the synchrotron knot located $\approx0.65^{\prime\prime}$ SE of the pulsar. The vectors included are as follows: spin-axis vector (SA) ($124\pm0.1^{\circ}$) (Ng \& Romani 2004), proper motion vector (PM) ($110\pm2\pm9^{\circ}$) (Kaplan et al. 2008), and the polarisation position angles of the pulsar ($105.97\pm2.00^{\circ}$) and synchrotron knot ($126.86\pm0.23^{\circ}$). Also, included is the INTEGRAL/IBIS measurement of the position angle during the off-pulse and bridge emission phases (OP+B) ($122.0\pm7.7^{\circ}$) (Forot et al. 2008).}
\label{figure4}
\end{center}
\end{figure}

\begin{table}[h]
\begin{center}
 \label{symbols}
 \begin{tabular}{@{}lccc}
  \hline
   & & Polarisation Degree (\%) & Position Angle ($^{\circ}$) \\
  \hline
$\gamma-ray$ (\textit{SPI} ) \cite{Dean08} 		    	 &OP  		&$46\pm10$  			&$123\pm11$\\
$\gamma-ray$ (\textit{IBIS}) \cite{Forot08} 		     	&OP  		&> 72  				&$120.6\pm8.5$\\
$\gamma-ray$ (\textit{IBIS}) \cite{Forot08} 		     	&OP+B  		&> 88  				&$122.0\pm7.7$\\
$\gamma-ray$ (\textit{IBIS}) \cite{Forot08} 		     	&P1 + P2 	&$42^{+30}_{-16}$ 		&$70\pm20$\\
$\gamma-ray$ (\textit{IBIS}) \cite{Forot08} 		    	&Total 		&$47^{+19}_{-13}$ 		&$100\pm11$\\
$\gamma-ray$ (\textit{IBIS}) (this work)     		     	&Total 		&$58\pm7$ 			&$85\pm10$\\
Optical (\textit{OPTIMA}) \cite{Aga09} 		      	&pulsar         &$9.8\pm0.1$ 	&$109.5\pm0.2$\\
Optical (\textit{HST/ACS}) (this work)				&pulsar 	&$4.90\pm0.33$ 		&$105.97\pm2.00$\\
Optical (\textit{HST/ACS}) (this work)				&knot 		&$61.70\pm0.72$ 		&$126.86\pm0.23$\\
X-ray (\textit{OSO 8}) (2.6--5.2 keV) \cite{Weisskopf78} 	&nebula         &$19.22\pm0.92$ 		&$155.79\pm1.37$\\
  \hline
 \end{tabular}
 \caption{List of multi-wavelength polarisation studies of the Crab Nebula and pulsar. For gamma-ray phase selections OP, B, P1, and P2 see text and \cite{Kuiper01}.}
\end{center}
\end{table}

\end{document}